\documentclass[twoside]{article}

\usepackage[arxiv]{aistats2022}

\usepackage[round]{natbib}

\usepackage[title]{appendix}
\usepackage{hyperref}
\usepackage{booktabs}
\usepackage{url}
\usepackage{amsmath}
\usepackage{amssymb}
\usepackage{array}
\usepackage{graphicx}
\usepackage[nameinlink]{cleveref}
\usepackage{algorithm}
\usepackage{algorithmicx}
\usepackage{algpseudocode}
\usepackage{listings}
\usepackage{xcolor}
\usepackage{courier}
\usepackage{makecell}
\usepackage{capt-of}

\newcommand{\bbR}{\mathbb{R}}
\newcommand{\bbE}{\mathbb{E}}

\newcommand{\calC}{\mathcal{C}}
\newcommand{\lnorm}{\left\lVert}
\newcommand{\rnorm}{\right\rVert}
\newcommand{\argmin}{\mathrm{argmin}}
\newcommand{\unif}{\mathrm{Unif}}
\newcommand{\argmax}{\mathrm{argmax}}
\crefname{appsection}{Appendix}{Appendices}

\algrenewcommand\algorithmicindent{0.5em}
\algnewcommand{\LineComment}[1]{\State \(\triangleright\) #1}
\algrenewcommand\algorithmicrequire{\textbf{Input:}}
\algrenewcommand\algorithmicensure{\textbf{Output:}}

\begin{document}

\twocolumn[

\aistatstitle{Focusing on Difficult Directions for Learning HMC Trajectory Lengths}

\aistatsauthor{ Pavel Sountsov \And Matt D. Hoffman }

\aistatsaddress{ Google Research } ]

\begin{abstract}
Hamiltonian Monte Carlo (HMC) is a premier Markov Chain Monte Carlo (MCMC) algorithm for continuous target distributions. Its full potential can only be unleashed when its problem-dependent hyperparameters are tuned well. The adaptation of one such hyperparameter, trajectory length ($\tau$), has been closely examined by many research programs with the No-U-Turn Sampler (NUTS) coming out as the preferred method in 2011. A decade later, the evolving hardware profile has lead to the proliferation of personal and cloud based SIMD hardware in the form of Graphics and Tensor Processing Units (GPUs, TPUs) which are hostile to certain algorithmic details of NUTS. This has opened up a hole in the MCMC toolkit for an algorithm that can learn $\tau$ while maintaining good hardware utilization. In this work we build on recent advances along this direction and introduce SNAPER-HMC, a SIMD-accelerator-friendly adaptive-MCMC scheme for learning $\tau$. The algorithm maximizes an upper bound on per-gradient effective sample size along an estimated principal component. We empirically show that SNAPER-HMC is stable when combined with mass-matrix adaptation, and is tolerant of certain pathological target distribution covariance spectra while providing excellent long and short run sampling efficiency. We provide a complete implementation for continuous multi-chain adaptive HMC combining trajectory learning with standard step-size and mass-matrix adaptation in one turnkey inference package.
\end{abstract}

\section{Introduction}

Hamiltonian Monte Carlo (HMC) \citep{duane1987hybrid, neal2011mcmc} is an efficient Markov Chain Monte Carlo (MCMC) algorithm for generating approximate samples from a continuous probability distribution $p(x)$ supported on $x \in \bbR^d$. For HMC, it is only necessary to be able to evaluate $p(x)$ up to a constant factor and to be able to compute its gradient. Such densities are commonly found as posteriors over parameters in the context of Bayesian inference, where the gradients can be obtained via automatic differentiation if the probabilistic model is implemented in a suitable programming language.

HMC generates proposals for a related distribution $p(x, p) \triangleq p(x) p(p)$ where $x$ and $p \in \bbR^d$ are identified as a fictitious ``position'' and ``momentum'' of a particle. $p(p)$ is typically chosen to be a zero-centered multivariate Gaussian with the density $p(p) \propto \exp(\frac{1}{2} p^T M^{-1} p)$ where $M$ is the covariance matrix, identified as the ``mass'' matrix, which is commonly chosen to be diagonal.

HMC generates a proposal in three steps. Given the current state $x_0$, a $p_0$ is sampled from $p(p)$. Then, Hamilton's equations: $\frac{\partial x}{\partial t} = \frac{\partial H}{\partial p}$, $\frac{\partial p}{\partial t} = -\frac{\partial H}{\partial x}$, where $H(x, p) \triangleq -\log p(x, p)$ is the Hamiltonian, are solved for the interval $t = [0, \tau]$ ($\tau$ is commonly known as the trajectory length). Typically an analytic solution is not available, so the dynamics are discretized and a suitable sympletic integrator, commonly the leapfrog integrator, is used with step size parameter $\epsilon$. The number of leapfrog steps $\ell$ is related to $\tau$ as $\ell = \lceil \tau / \epsilon \rceil$. The proposed state $(x_\tau, -p_\tau)$ (the momentum is negated to ensure a reversible proposal) is accepted or rejected using the Metropolis-Hastings step \citep{metropolis1953equation, hastings1970mh} that targets the joint $p(x, p)$ distribution.

The efficiency of the HMC sampler strongly depends on the settings of its hyperparameters $\tau$, $\epsilon$ and $M$. $M$ and $\epsilon$ are typically set either continuously via Adaptive MCMC \citep{andrieu2008tutorial} or windowed adaptation popularized by \citet{carpenter2017stan}. In both cases, $\epsilon$ is adjusted such that the Metropolis-Hastings acceptance probability is between 0.6 and 0.9 \citep{betancourt2014optimizing}. $M$ is typically chosen to be a diagonal matrix with entries set to sample variances. How to set $\tau$ well is less obvious. \citet{hoffman2011no} proposed a variant of HMC called the No-U-Turn Sampler (NUTS) that dynamically chooses $\tau$ without breaking detailed balance. This allows it to adapt to the local geometry of the distribution, although the requirement for detailed balance requires the algorithm to discard on average half of considered states when generating a proposal. This combination of adaptive methods is the standard way of using HMC, but it is not well adapted to the modern modern SIMD-based accelerators. A core issue is that those accelerators prefer to evaluate multiple chains in parallel, and when using NUTS, different chains may choose different numbers of leapfrog steps. SIMD implementations of NUTS exist \citep{radul2019automatically, phan2019iterative, lao2019unrolled} but all suffer from significant wall-clock and utilization overhead compared to a non-dynamic HMC  \citep{radul2019automatically,hoffman2021adaptive}.

Alternate ways of setting $\tau$ are based on the Adaptive MCMC framework, e.g. Adaptive HMC \citep{wang2013adaptive}, eHMC \citep{wu2018faster} and ChEES-HMC \citep{hoffman2021adaptive}. They function by defining a criterion $\calC(\tau)$ as proxy for sampler efficiency and adjusting $\tau$ to maximize it. Being proxies, they can sometimes provide inaccurate estimates of the optimal $\tau$. In this work, we point out certain shortcomings of the ChEES criterion, the current best choice amongst criteria in this setting, and propose ways of solving them. We do so by defining a new criterion, SNAPER, and make the following contributions:

\begin{enumerate}
    \item We show how SNAPER is robust to pathological eigenspectra of the covariance of $x$ as well as instabilities in stochastic optimization of $\tau$.
    \item We show how to incorporate SNAPER into a unified, windowless, continuous adaptation recipe to learn the main hyperparameters of HMC.
    \item We show how SNAPER combined with the adaptation recipe produces a sampler has high sampling efficiency as well as requiring a low total number of gradient evaluations (including warmup period).
\end{enumerate}

\section{Related Work}

In this section, we review the existing criteria used as proxies for HMC sampler efficiency that are then used to tune the $\tau$ hyperparameter. A typical use case of MCMC is to estimate expectations of a test function $f(x)$, i.e. $\theta_f \triangleq \bbE[f(x)]$. The MCMC estimator of $\theta_f$ can be constructed as $\hat{\theta}_f \triangleq \frac{1}{n} \sum_{t=0}^n f(x^{(t)})$ where $x^{(t)}$ is the $t$'th iteration of the MCMC chain. When the MCMC central limit theorem applies, the variance of this estimator can be computed as $Var[\hat{\theta}_f] = \frac{Var[\theta_f]}{ESS_f}$ where $ESS_f$ is the Effective Sample Size. A convenient way of expressing ESS is in terms of the autocorrelation function $\rho_f(l) \triangleq \frac{\bbE[((f(x^{(t)}) - \theta_f)(f(x^{(t + l)}) - \theta_f]}{Var[f(x)]}$:
\begin{align}
    \label{eq:ess}
    ESS_f \triangleq \frac{n}{1 + 2 \sum_{l=1}^{\infty} \rho_f(l)}.
\end{align}
$ESS/grad$ is a typical measure of efficiency, where $grad$ is the number of gradient evaluations needed to compute the estimator. It is typical to assume that the gradient evaluations dominate the wallclock runtime of algorithm. For HMC using the standard leapfrog integrator the number of gradient evaluations is $\ell$ in one iteration.

ESS requires long MCMC trajectories to estimate accurately which makes it impractical to use as a criterion for online hyperparameter learning. Instead most adaptive HMC algorithms optimize some variant of Expected Square Jump Distance (ESJD), defined as $\bbE \left[ \lnorm x^{(0)} - x^{(t + 1)}(\tau)\rnorm_2^2 \right]$, which only requires considering two adjacent iterations. eHMC \citep{wu2018faster} and Adaptive HMC\citep{wang2013adaptive} optimize ESJD and and $\mathrm{ESJD} / \ell^{\,0.5}$ respectively. The U-Turn criterion from NUTS is also derived from ESJD. \citep{hoffman2021adaptive} show that optimizing for ESJD produces overlong trajectory lengths, and instead propose the ChEES criterion which can be expressed by first transforming $x$ with:
\begin{align}
  \label{eq:f_chees}
  f_{\mathrm{ChEES}}(x) = \frac{1}{2}\lnorm z \rnorm_2^2,
\end{align}
where $z \triangleq x - \bbE[x]$, before computing the generalized ESJD:
\begin{align}
  \label{eq:esjd}
  ESJD_f = \bbE \left[ \lnorm f(x^{(t)}) - f(x^{(t + 1)}(\tau)) \rnorm_2^2 \right].
\end{align}

An alternate research program explored the use of proposal entropy to optimize the proposal hyperparameters \citep{titsias2019gradient,li2021neural}. This method relies on having a tractable distribution $q(x^{(t + 1)} | x^{(t)})$ which is available for Random-Walk Metropolis (RWM) and Metropolis Adjusted Langevin Algorithm (MALA), but is notably not available for HMC.

\section{Shortcomings of the ChEES Criterion}

\begin{figure*}[!tb]
  \centering
    \includegraphics[scale=0.55]{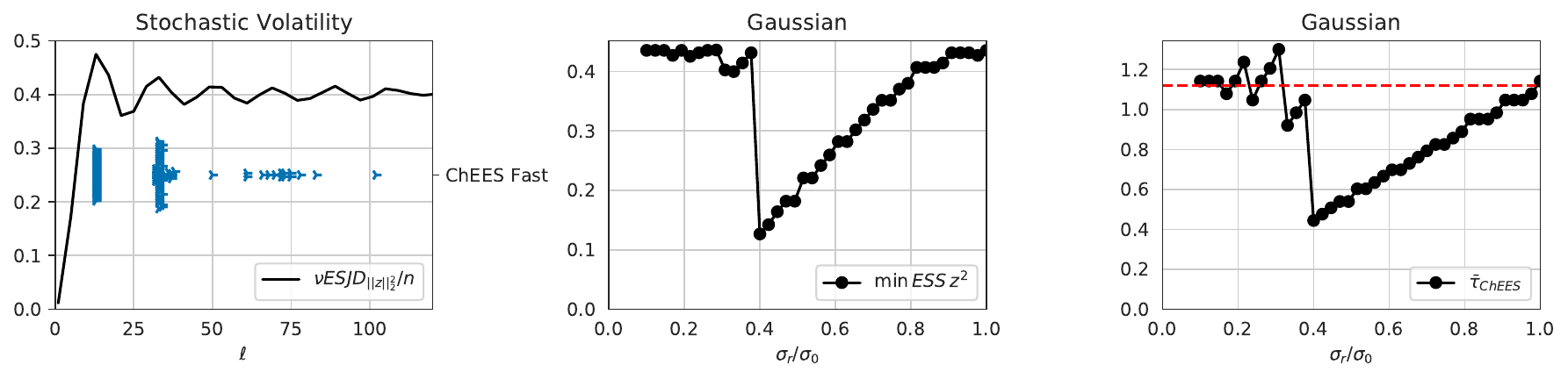}
\caption{
    Shortcomings of the ChEES criterion. The non-decay of the 
    criterion with increasing number of leapfrog steps causes 
    gradient-based adaptation to sometimes overshoot the global optimum 
    (left). The ChEES-found number of leapfrog steps across 100 runs 
    are displayed as jittered points, with jitter amount proportional 
    to local point density. For target distributions where the 
    covariance has a large number of medium-sized eigenvalues (with 
    magnitude $\sigma_r$) and few large eigenvalues ($\sigma_0$), 
    ChEES-HMC finds trajectory lengths smaller than optimal (dashed red line) (right), 
    causing a drop in ESS (middle). The jagged nature of the right plot arises from optimum jumping between peaks (similar to ones in the left panel) as they shift.
}
    \label{fig:motivation}
\end{figure*}

In this work, we focus our analysis on the ChEES criterion. While it has been shown to be a very competitive criterion across a number of test functions and benchmark problems, it has number of shortcomings as a criterion, which we illustrate in \Cref{fig:motivation}. First, it does not incorporate the cost of generating a proposal. Consider the case where we jitter $\tau$ by sampling it from $\unif(0, 2\bar{\tau})$ before every HMC step. Jittering the trajectory length is a recommended practice to minimize resonances \citep{neal2011mcmc}.  Left panel of \Cref{fig:motivation} plots ChEES versus the mean number of leapfrog steps $\bar{\ell} = \lceil \bar{\tau} / \epsilon \rceil$ as well as the values of $\bar{\ell}$ found by ChEES-HMC (100 separate runs using the ``Fast'' configuration described below). While some trials find the first peak of the criterion (corresponding to the global optimum for this problem), stochastic optimization can easily overshoot and find overlong values for $\bar{\ell}$. This can be mitigated by the use of a more gentle learning rate at the cost of adaptation speed. This problem also exists, in principle, with regular ESJD that is used by NUTS, but there the algorithm is constructed to stop at the first peak it finds, which is not an easy thing to do with stochastic optimization. This issue is exacerbated if the $p(x, p)$ changes over the course of adaptation when $M$ is co-adapted.

The second shortcoming is that ChEES can be fooled by certain eigenspectra of the covariance matrix of $x$ to prefer a trajectory length that is too short. Consider a zero-centered multivariate Gaussian $p(x)$ with a diagonal covariance matrix set to $\{\sigma_0^2, \sigma_r^2 ... \sigma_r^2\}$, with the trailing $\sigma_r^2$ repeated 300 times and $\sigma_0 \ge \sigma_r$. We measure $\min ESS_{z^2}$, the minimum ESS across components of $z^2$, which, for this problem, is maximized by $\hat{\tau} \approx 1.12 \sigma_0$ (see Section 2.1 of \cite{hoffman2021adaptive}).

As we vary $\sigma_r / \sigma_0$ we compute the mean trajectory length $\bar{\tau}^*_{\mathrm{ChEES}}$ that maximizes ChEES (\Cref{fig:motivation}, right panel) and $\min ESS_{z^2}$ evaluated at $\bar{\tau}^*_{\mathrm{ChEES}}$ (\Cref{fig:motivation}, middle panel). While ChEES's optimum matches the optimal trajectory length when the $\sigma_r$ is either very small (meaning they barely affect the value of ChEES) or very large (meaning that $p(x)$ is close to being isotropic), it struggles in the intermediate regime. There the repeated small eigenvalues overwhelm the contribution to ChEES from $\sigma_0$, causing ChEES to prefer shorter-than-optimal trajectories. This is caused by ChEES computing an average of per-dimension autocorrelation sums, each weighted by $\sigma_i^4$. While it heavily weights directions with greatest variance, the weighting can be overcome as illustrated here.

\section{SNAPER Criterion}
\label{sec:snaper}

In this section we motive a new criterion to address the issues of ChEES criterion: its insensitivity to proposal cost and its dependence on eigenspectrum of covariance of $x$. A natural way to incorporate a notion of cost into ChEES is to normalize it by some increasing function of $\ell$. \cite{wang2013adaptive} optimize $ESJD / \ell^{\,0.5}$ with a heuristic motivation. In practice, it is more convenient to use $\tau$ in place of $\ell$, to avoid issues with taking gradients with respect a discrete value. We thus define the ChEES-Rate (ChEESR) criterion:
\begin{align*}
    ChEESR \triangleq \frac{ESJD_{\lnorm z \rnorm_2^2}}{\tau}
\end{align*}
What does ChEESR optimize? When the chain is stationary we can show that $ESJD_f = Var [f(x)](1 - \rho_f(1))$. This identity has been used to motivate ESJD as a criterion \citep{pasarica2010adaptively}, as $ESS_f$ is related to $\rho_f(1)$ via a decreasing nonlinear function (\Cref{eq:ess}). It's not obvious that this connection allows using $ESJD_f / \tau$ as a proxy for $ESS_f / grad$ however. To better motivate this, we recall that, given a square integrable $f(x)$ and a reversible MCMC proposal, it is possible to bound $ESS$ from above (see \Cref{sec:ess_max}) by:
\begin{align}
\label{eq:ess_max}
    ESS_f \le n \frac{1 - \rho_f(1)}{1 + \rho_f(1)} \triangleq ESS^{\max}_f.
\end{align}
This bound is the equivalent to assuming $\rho_f(l) = \rho_f(0)^l$ and substituting that into \Cref{eq:ess}. If we assume positive $\rho_f(1)$ then $ESJD_f$ is an upper bound (up to a constant factor) on $ESS$ and therefore ChEESR is proportional to an upper bound on $ESS_{\lnorm z \rnorm_2^2} / \tau$.

When doing Bayesian inference it is common to desire to maximize the minimum componentwise $ESS$, e.g. $\min ESS_{z^2}$. Starting from ESJD there are three steps of approximation between it and $\min ESS_{z^2}$: $\nu ESJD_f \ge_1 ESS^{\max}_f \ge_2 ESS_f \approx_3 \min ESS_{z^2}$ where $\nu ESJD_f \triangleq \frac{n}{Var[f(x)]} ESJD$. We cannot control the first inequality except to hope that $\rho_f(1)$ is small. The second inequality can be measured empirically with minimal memory overhead if $f(x)$ is scalar, which it is for $f_{\mathrm{ChEES}}$, as we can use \Cref{eq:ess} directly. The tightness of that bound can be recruited as a post-hoc certificate that criterion optimization likely found a good optimum, although a practitioner would not have much recourse to fix the optimization if the bound is loose outside of adjusting the parameterization of the model. The third approximation occasionally has pathologies when using $f_{\mathrm{ChEES}}$ as shown by the example in \Cref{fig:motivation}. We propose adjusting $f(x)$ to resolve that failure case.

Identifying $\argmin_i ESS_{z_i^2}$ is difficult, but a reasonable heuristic is to focus on optimizing ESS along the direction of greatest variance. This has been part of the motivation behind ESJD and ChEES as they weigh directions by $\sigma_i^2$ and $\sigma_i^4$ respectively. Instead of doing this implicit weighting, we propose identifying the direction of greatest variance explicitly by identifying it with the first principal component of the $z$ covariance matrix and then learning it via Principal Component Analysis. A convenient algorithm in this field is Oja's method \citep{oja1982simplified} which is memoryless and is easy to integrate into the rest of the adaptive MCMC framework. In practice, we use a minibatch version of the algorithm as we have access to the state of all the chains at once (\Cref{alg:oja}). In summary, we propose to maximize the Squared Norm Along Principal component ESJD Rate (SNAPER):
\begin{align}
    \label{eq:snaper}
    SNAPER &\triangleq \frac{ESJD_{(z^Tw)^2}}{\tau} \\
    f_{\mathrm{SNAPER}}(x) &= (z^Tw)^2
\end{align}
where $w$ is the learned first principal component.

\section{SGA-HMC}

\begin{figure*}[!tb]
  \centering
    \includegraphics[scale=0.55]{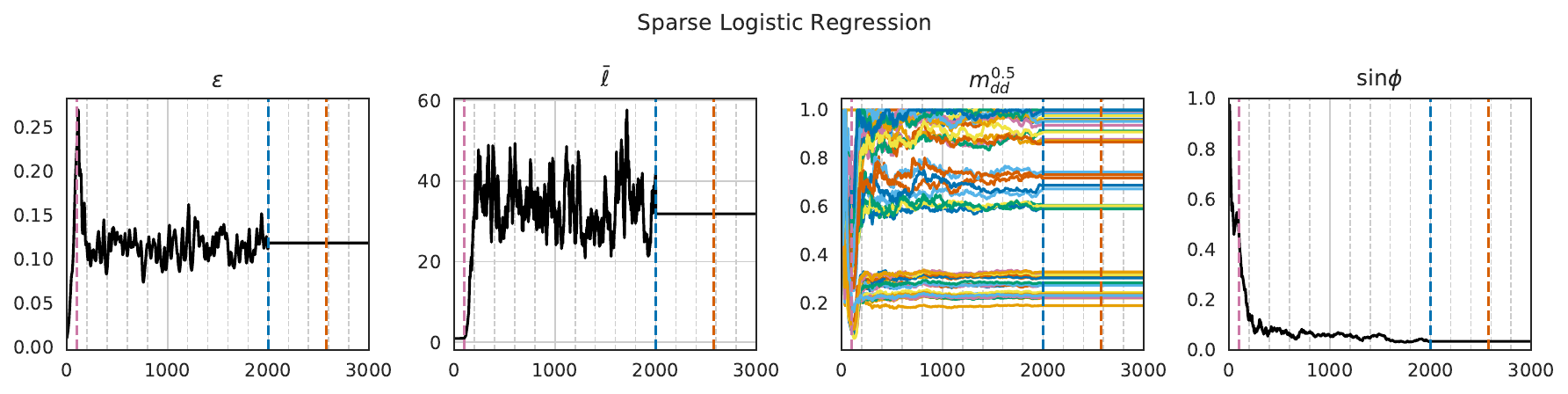}
\caption{
    Evolution of the learned hyperparameters from a SNAPER-HMC run on the German Credit Sparse Regression problem as a function of MCMC iteration number. During the initial phase (first vertical line), the number of leapfrog steps is set to 1. Then, the adaptation proceeds for a chosen number of steps (second vertical line), after which the hyperparameters are fixed. Sampling can be stopped when $\hat{R} < 1.01$ (third vertical line). The last panel is the sin of the angle between the learned principal component direction and its projection onto the top 15 eigenvectors of the unconstrained state covariance matrix whose scales are greater than 0.85 of the largest scale.
}
    \label{fig:trace}
\end{figure*}

In this section we discuss how to incorporate the SNAPER criterion into a specific instance of an adaptive MCMC algorithm. Instead of more typical windowed adaptation setup popularized by \citet{carpenter2017stan} we propose to adapt the hyperparameters continously with a decaying adaptation rate. See \Cref{sec:sga_hmc_details} for additional details. We adjust $\epsilon$ using ADAM \citep{kingma2015adam}, with a synthetic gradient derived from the difference between the harmonic mean of the acceptance probability and a target acceptance probability of $\alpha^* = 0.8$. To ensure a reasonable value for $\epsilon$ early in the adaptation period, we fix $\ell = 1$ for the first 100 iterations. $M$ is learned using a variant of Welford's algorithm \citep{welford1962note} which uses an exponentially weighted average with a decaying learning rate to smoothly discount early, biased iterates. The same algorithm is also used for centering $z$ when computing the value of the SNAPER, ChEES and ChEESR criteria. $\tau$ is taken to be sampled from $\unif(0, 2\bar{\tau})$, where $\bar{\tau}$ is learned using ADAM following the pathwise gradients computed via automatic differentiation. For SNAPER, we learn $w$ using the minibatch Oja's algorithm \Cref{alg:oja}. The pseudocode for the combined algorithm, Stochastic Gradient Ascent HMC (SGA-HMC) is listed in \Cref{alg:sga_hmc}\footnote{Implementation available at \url{https://github.com/tensorflow/probability/tree/main/discussion/snaper_hmc}}. While adaptation is proceeding, we maintain iterate averages of all the hyperparameters for use in post-warmup sampling. \Cref{fig:trace} illustrates the action of SGA-HMC with the SNAPER criterion on one of the example problems.

As a baseline, we also define an Adaptive NUTS algorithm, which follows \Cref{alg:sga_hmc} except that HMC is replaced by NUTS. Critically, for the first 100 steps we still only take one leapfrog step as with SGA-HMC, to make sure NUTS starts out with a reasonable $\epsilon$.

\begin{algorithm}
\caption{Stochastic Gradient Ascent HMC}\label{alg:sga_hmc}
\begin{algorithmic}[1]
    \Require $x^{(t, k)}$, an a minibatch of d-dimensional states indexed by batch index $k \in [1, b]$. $\bar{\tau}^{(t)}$, the mean trajectory length. $\epsilon^{(t)}$, the step size. $\mu^{(t)}, \sigma^{2(t)}$, running first and second moments. $\eta_{\bar{\tau}}$, $\eta_\epsilon$, learning rate for $\bar{\tau}$ and $\epsilon$ hyperparameters. $t$, the iteration index.
    \Ensure Next state $x^{(t+1)}$. Updated versions of parameters and statistics.
    
    \LineComment{Compute current HMC hyperparameters.}
    \State $\tau^{(t)} \sim \unif(0, 2 \bar{\tau}^{(t)})$
    \State $m^{(t)}_{ii} \leftarrow \sigma^{2(t)}_i / \max_i \sigma^{2(t)}_i$
    \LineComment{Clip $\tau^{(t)}$ during the first 100 iterations.}
    \If{$t < 100$}
        \State $\tau^{(t)} \leftarrow \epsilon^{(t)}$ 
    \EndIf
    \LineComment{Generate the next state using HMC.}
    \State $x^{(t + 1)}, x'^{(t)}, \alpha^{(t)} \leftarrow \textsc{HMC}(x^{(t)}, \epsilon^{(t)}, \tau^{(t)}, M^{(t)}$)
    \LineComment{Update trajectory distribution}.
    \State $\calC^{(t)} \leftarrow \calC(x^{(t)}, x'^{(t)}, \alpha^{(t)}, \mu^{(t)})$
    \State $\log \bar{\tau}^{(t + 1)} \leftarrow \textsc{ADAM}(\log \bar{\tau}^{(t)}, -\frac{d \calC^{(t)}}{d \log \bar{\tau}^{(t)}}, \eta_{\bar{\tau}}, \beta_1, \beta_2)$
    \LineComment{Update step size.}
    \State $\delta_\epsilon \leftarrow \alpha^* - (\frac{1}{b} \sum_k \alpha_k^{-1})^{-1}$
    \State $\log \epsilon^{(t + 1)} \leftarrow \textsc{ADAM}(\log \epsilon^{(t)}, \delta_\epsilon, \eta_\epsilon)$
    \LineComment{Update mean and variance estimates.}
    \State $\eta_\mu \leftarrow \frac{1}{\lceil t / \kappa \rceil + 1}$
    \State $\mu^{(t + 1)} \leftarrow (1 - \eta_\mu)\mu^{(t)} + \frac{\eta_\mu}{b} \sum_k x^{(t + 1, k)}$
    \State $\sigma^{2(t + 1)} \leftarrow (1 - \eta_\mu)\sigma^{2{(t)}} + \frac{\eta_\mu}{b} \sum_k (x^{(t + 1, k)} - \mu^{(t)})^2$
\end{algorithmic}
\end{algorithm}

\section{Experiments}

We validate SNAPER and ChEESR performance on six problems, three applied to real world data and three applied to synthetic data sampled from the respective model's prior. All the models are taken from the Inference Gym \citep{sountsov2020inference}, which we refer the reader to for model definitions. We summarize the models used in \Cref{tab:models}. For each model, we run SGA-HMC with one of the criteria for a prescribed number of adaptation steps, we then turn off adaptation, switch HMC to use the iterate-averaged hyperparameters and collect statistics like ESS/grad and potential scale reduction \citep{gelman1992,vehtari2021rank}. Some of the models have constraints on their parameters, which we handle by reparameterizing them and adjusting the log-density by the logarithm of the determinant of the Jacobian. SGA-HMC is ran and all of the statistics are computed in the unconstrained space. For each run, we initialize the chains at the prior mean. This severely underdispersed initialization aids algorithms which rely on cross-chain statistics, although it leaves us vulnerable to not detecting all the modes of the posterior. We consider SNAPER, ChEESR and ChEES criteria. For ChEES we considered two settings of the trajectory length learning rate $\eta_{\bar{\tau}} = [0.025, 0.05]$, the latter being identified in the plots and tables as ``ChEES fast''. For all other conditions $\eta_{\bar{\tau}}$ was set to $0.05$. See \Cref{sec:hyperparameters} for more hyperparameter settings.

\begin{table*}[!tb]

\centering
\begin{tabular}{ lll }
\toprule
 Model & $d$ & Inference Gym Snippet \\
\midrule
Logistic Regression & 25 & \lstinline[]$GermanCreditNumericLogisticRegression()$ \\
Sparse Logistic Regression & 51 & \makecell[l]{\lstinline[]$GermanCreditNumericSparseLogisticRegression($ \\ \hspace{0.25in} \lstinline[]$positive_constraint_fn="softplus")$} \\
Brownian Bridge & 32 & \makecell[l]{\lstinline[]$BrownianMotionUnknownScalesMissingMiddleObservations($ \\ \hspace{0.25in} \lstinline[]$use_markov_chain=True)$} \\
Item Response Theory & 501 & \lstinline[]$SyntheticItemResponseTheory()$ \\
Hierarchical Linear Model & 97 & \lstinline[]$RadonContextualEffectsIndiana()$ \\
Stochastic Volatility & 2519 & \lstinline[]$VectorizedStochasticVolatilitySP500()$ \\
\bottomrule
\end{tabular}

\caption{
Bayesian posteriors used for the experiments. The snippets refer to the Python class invocation to construct the model (the classes are all found in the \lstinline{inference_gym.targets} module).
}
\label{tab:models}

\end{table*}

\subsection{Long Run Efficiency}

\begin{figure*}[!tb]
  \centering
    \includegraphics[scale=0.55]{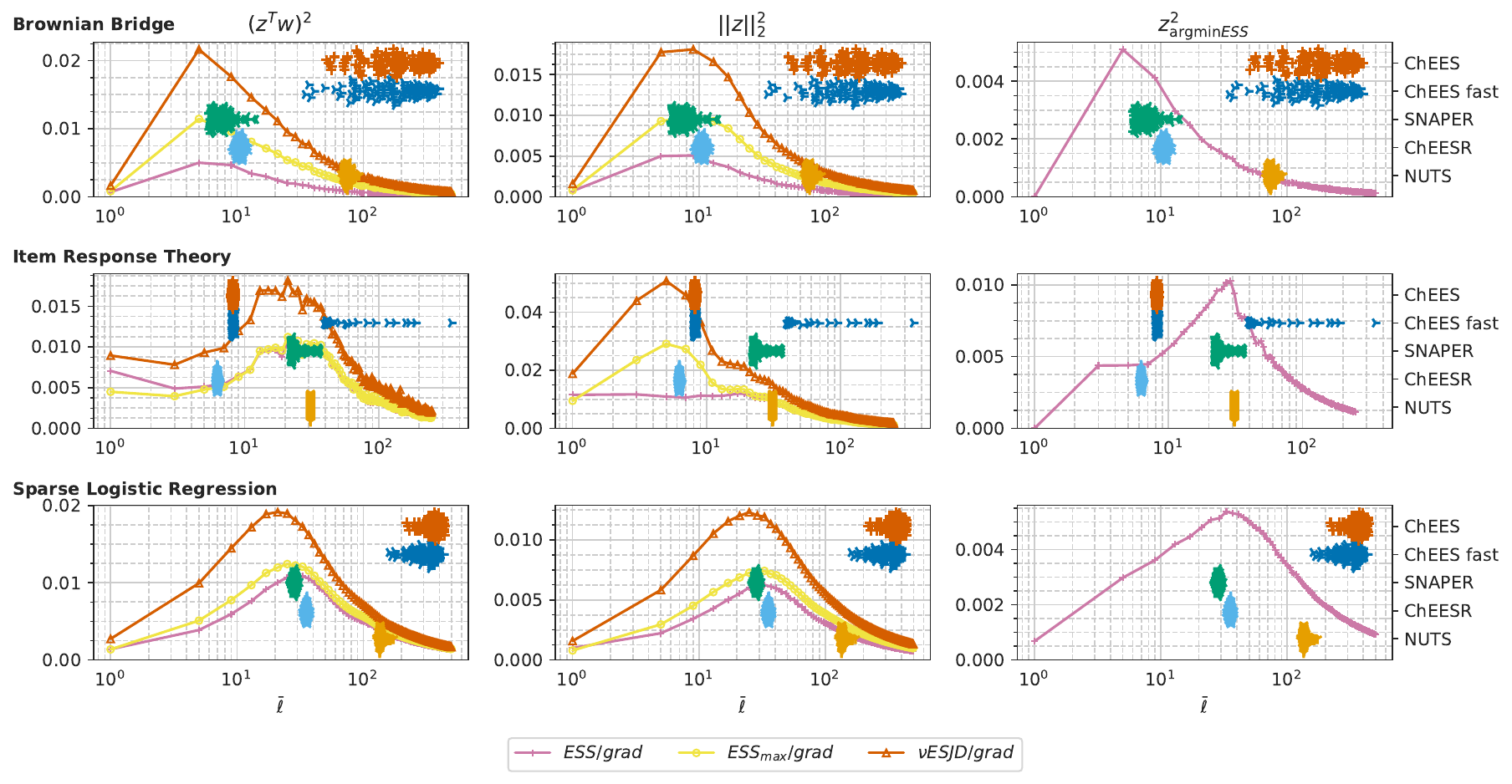}
\caption{
    Long run behavior of different criteria. The rows are three 
    representative target distributions. The columns are three test 
    functions: squared projected norm, squared norm and squared 
    component with the lowest ESS/grad. The lines are computed via 
    sweeps over the number of mean leapfrogs per MCMC iteration (higher is better). The 
    points are jittered final mean numbers of leapfrog steps across 100 
    runs, with jitter amount proportional to local point density. 
    SNAPER tends to find a good trajectory length as measured by all 
    ESS/grad metrics.
}
    \label{fig:asymptotics}
\end{figure*}

For the first experiment, we examined the long run sampling efficiency of all the algorithms. This is the setting where we can discount the cost of the warmup iterations and are solely interested in collecting a large number of samples from long chains. First, we performed sweeps of SGA-HMC with fixed $\bar{\ell}$ and $M$ and measured $ESS/grad$, $ESS^{\max}/grad$ and $\nu ESJD/grad$ for three test functions, $f_{\mathrm{SNAPER}}$, $f_{\mathrm{ChEES}}$ and $z^2_{argmin ESS}$ (lines in \Cref{fig:asymptotics}). For Item Response Theory and Sparse Logistic Regression $ESS^{\max}$ is close to actual $ESS$ when using $f_{\mathrm{SNAPER}}$ (left column). Let $\bar{\ell}^*_{\calC} \triangleq \argmax_{\bar{\ell}} \calC(\ell)$. For these models, $\bar{\ell}^*_{\mathrm{SNAPER}}$ is close to $\bar{\ell}^*_{ ESS/grad}$. The bound is not as tight for Sparse Logistic Regression when using $f_{\mathrm{ChEES}}$, middle column, or Brownian Bridge, which causes the performance to be less predictable. We reiterate that the bound of $ESS^{\max}/grad \ge ESS/grad$ can be computed empirically with minimal memory overhead, because both $f_{\mathrm{ChEES}}$ and $f_{\mathrm{SNAPER}}$ are scalar functions. This makes it accessible as a diagnostic for ChEES and SNAPER-HMC. An example of a pathological covariance spectrum similar to \Cref{fig:motivation} is the Item Response Theory model. As expected, we see that $\bar{\ell}^*_{\mathrm{SNAPER}}$ matches $\bar{\ell}^*_{\min ESS_{z^2} / grad}$ closely, while $\bar{\ell}^*_{\mathrm{ChEESR}}$ is too short for this target distribution.

We then ran 100 separate trials of SGA-HMC targeting each criterion, each trial being distinguished by the PRNG seed used for the random sampling events. We warm up for 5000 steps, followed by 1000 steps of sample collection. We see that both ChEESR and SNAPER recover the optima of their respective criteria reliably. ChEES and ChEES-fast have a lot more variability which is shown by both the large spread of learned trajectories for Brownian Bridge (note the log-scale on the x-axis) and the occasional overshoot for Item Response Theory. In general, ChEES-derived metrics either find trajectories that are too long or too short, depending on the problem, causing suboptimal $\min ESS_{z^2}$. SNAPER results in consistently good performance across all problems (especially when $ESS^{\max}$ bound is tight). Additional results can be seen in \Cref{tab:asymptotics_tabl} and \Cref{sec:extra_plots}.

\subsection{Short Run Efficiency}

\begin{figure*}[!tb]
  \centering
    \includegraphics[scale=0.55]{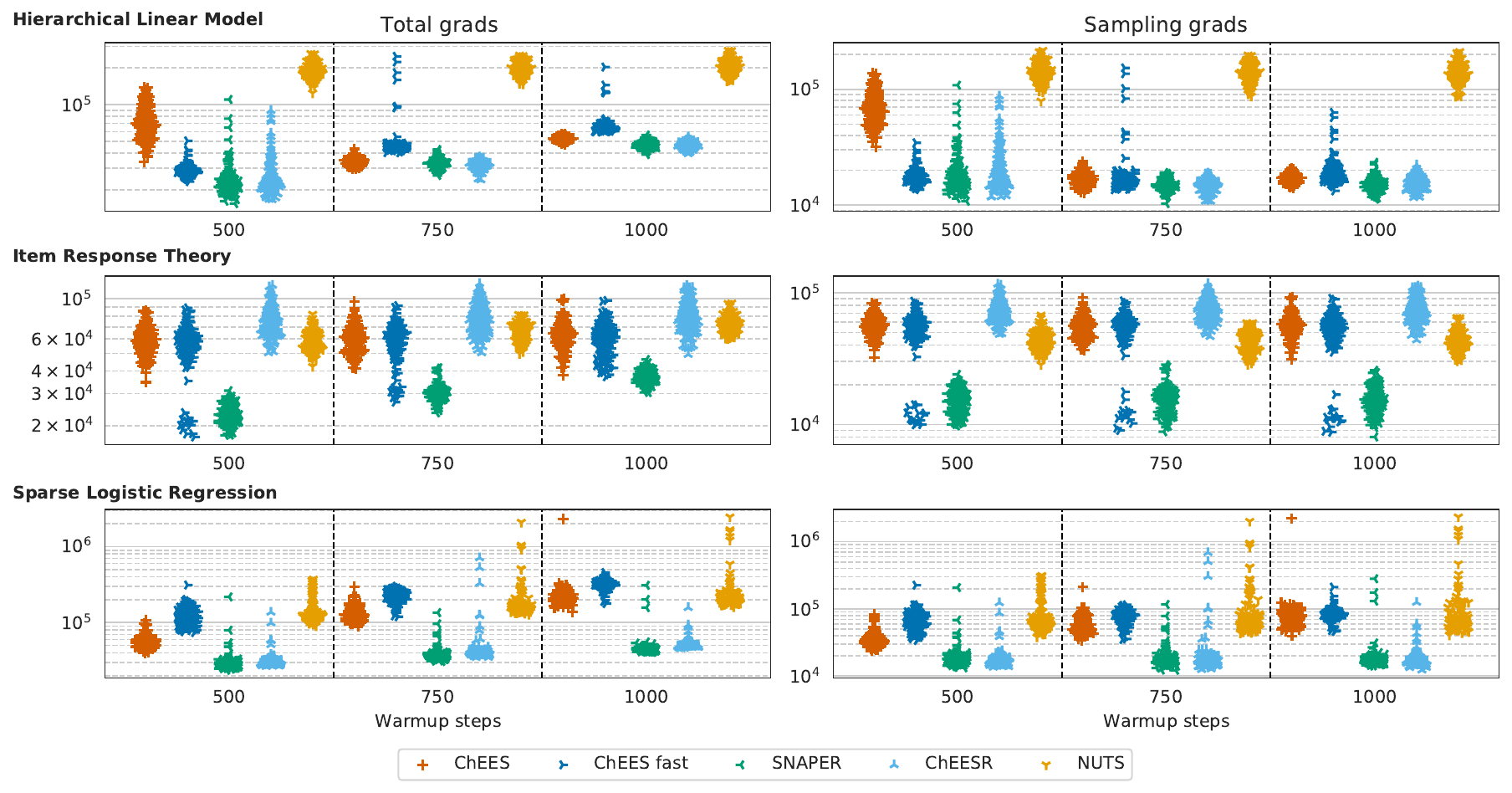}
\caption{
    Gradient evaluation counts as a function of number of warmup steps. 
    The three rows are three representative target distributions. The left 
    column counts the amount of grads for both warmup and sampling 
    periods, while the right counts sampling grads only. Sampling stops 
    when the post-warmup chains reach an $\hat{R} < 1.01$. Each setting 
    of the warmup steps represents a separate run from the other 
    settings.
}
    \label{fig:early_stop}
\end{figure*}

Many questions in Bayesian inference only require less than 1000 independent samples, sometimes as low as 100. If we run 100 chains in parallel, then in principle we could stop and record just the final iteration as soon as we had a certificate that the sampler has converged. $\hat{R}$ is a popular convergence diagnostic that approaches 1 from above as the chain converges to its stationary distribution; $\hat{R} < 1.01$ is considered an acceptable threshold for convergence \citep[e.g.]{vats2018revisting}. Thus, we envision the following protocol: 1) Run SGA-HMC for a prescribed number of warmup/adaptation steps. 2) Turn off adaptation and collect samples until $\hat{R} < 1.01$ for all test functions of choice (in this experiment, we use $f(x) = x_i$ for all $i \in [1, d]$). A good algorithm will reach this stopping condition with few gradient evaluations. There is a non-zero optimal number of warmup steps for this procedure: too few, and the sampling stage will proceed too slowly due to a poorly adapted sampler; too many, and the adaptation time is wasted. In \Cref{fig:early_stop} we run SGA-HMC for all four settings, and stop adaptation at 500, 750 and 1000 steps. We run 100 separate trials for each experiment configuration. We report both the total number of gradient evaluations (left column) and the number of gradient evaluations just in the sampling phase (right column). We observe that ChEES-fast sometimes will overshoot the optimum (Hierarchical Linear Model, 1000 warmup steps). This deficiency can be ameliorated with a slower learning rate, but at a cost reduced performance in some settings (Hierarchical Linear Model, 500 warmup steps). In general, for ChEES and ChEES-fast, stopping early can recover some of the competitiveness with other criteria because $\bar{\tau}$ has not yet converged to its overlong critical point. ChEESR and SNAPER behave similarly to each other except in the case of Item Response Theory where the pathological covariance spectrum causes ChEESR to learn an inefficient sampler. Additional results are available in \Cref{tab:early_stop_table} and the \mbox{\Cref{sec:extra_plots}}. When ChEESR can learn a good sampler, it tends to behave well in these experiments. SNAPER sometimes struggles with very short warmup windows, perhaps due to the additional noise caused by needing to learn the value of the first principal component.

\begin{table*}[!tb]

\centering

\begin{tabular}{ lllllll }
\toprule
 & Grid & ChEES & ChEES fast & SNAPER & ChEESR & NUTS \\
\midrule
Brownian Bridge & 4.71e-3 & 1.30e-4 & 1.30e-4 & \textbf{2.97e-3} & 2.52e-3 & 3.84e-4 \\
Hierarchical Linear Model & 5.99e-3 & 4.67e-3 & 4.63e-3 & 5.18e-3 & \textbf{5.41e-3} & 1.28e-3 \\
Item Response Theory & 1.02e-2 & 3.42e-3 & 3.36e-3 & \textbf{8.05e-3} & 2.61e-3 & 4.37e-3 \\
Logistic Regression & 1.20e-1 & 9.61e-2 & 9.76e-2 & 1.16e-1 & \textbf{1.32e-1} & 8.94e-3 \\
Sparse Logistic Regression & 5.37e-3 & 1.10e-3 & 1.18e-3 & 4.98e-3 & \textbf{5.21e-3} & 1.44e-3 \\
Stochastic Volatility & 1.37e-2 & 1.44e-2 & 3.31e-3 & \textbf{1.49e-2} & 1.27e-2 & 5.14e-3 \\
\bottomrule
\end{tabular}

\caption{
Asymptotic $\min ESS_{z^2} /grad$ per experiment configuration (including grid search), higher is better. The numbers 
are 10'th percentile across 100 runs. Bold counts are the best 
configurations for each problem (not counting the grid search).
}
\label{tab:asymptotics_tabl}

\end{table*}

\begin{table*}[!tb]

\centering

\begin{tabular}{ llllll }
\toprule
 & ChEES & ChEES fast & SNAPER & ChEESR & NUTS \\
\midrule
Brownian Bridge & 2.7e+5/2.9e+5 & 4.0e+5/7.1e+5 & 8.7e+4/8.3e+4 & \textbf{7.8e+4}/7.9e+4 & 2.8e+5/3.8e+5 \\
Hier. Linear Model & 1.2e+5/7.1e+4 & 5.8e+4/7.3e+4 & \textbf{4.7e+4}/5.8e+4 & 5.9e+4/5.7e+4 & 2.1e+5/2.1e+5 \\
Item Response & 7.4e+4/7.7e+4 & 7.6e+4/7.5e+4 & \textbf{3.3e+4}/4.0e+4 & 9.7e+4/1.0e+5 & 6.8e+4/7.6e+4 \\
Logistic Reg. & 2.9e+3/3.9e+3 & 2.9e+3/3.9e+3 & \textbf{2.8e+3}/3.7e+3 & 2.9e+3/3.7e+3 & 3.6e+4/4.9e+4 \\
Sparse Logistic Reg. & 1.3e+5/3.5e+5 & 3.0e+5/4.1e+5 & \textbf{4.0e+4}/4.9e+4 & 4.3e+4/5.6e+4 & 2.6e+5/3.0e+5 \\
Stochastic Volatility & 2.1e+4/2.2e+4 & 2.0e+4/2.2e+4 & 1.8e+4/1.8e+4 & 1.7e+4/\textbf{1.7e+4} & 5.5e+4/6.8e+4 \\
\bottomrule
\end{tabular}

\caption{
Total count of gradient evaluations, lower is better. The two 
counts are the 90'th percentile across 100 runs for 500 and 750 warmup 
steps respectively. Bold counts are the best configurations for each 
problem.
}
\label{tab:early_stop_table}

\end{table*}
\section{Discussion}

In this work we have introduced a new criterion for gradient-based HMC trajectory length adaptation -- SNAPER. We have illustrated that SNAPER has excellent long and short chain efficiency. In most cases, with sufficient warmup, SNAPER delivers consistently optimal trajectory lengths. SNAPER derives these quantities by having an accurate cost model of HMC sampling efficiency and by using a good heuristic for identifying which directions to focus on when optimizing $\tau$. An epiphenomenon of this robustness is that the criterion tends to have a single, well defined peak (compare \Cref{fig:motivation}, left column, and \Cref{fig:asymptotics}, left column) allowing the use of high learning rates.

When does SNAPER-HMC fail? SNAPER-HMC relies on the principal component identifying the most ``difficult'' direction, but it's easy to come up with a counterexample where a low variance direction has enough nonlinear twists to require a larger $\tau$ to explore well than simpler, high variance dimensions might suggest. If these difficult dimensions can be identified manually, then a practitioner could hand-craft a $w$ (\Cref{eq:snaper}) that focuses on those dimensions as a remedy to this problem. A final limitation is that since SNAPER is relies on the quality of the $ESS^{\max}$ bound, when that bound is loose SNAPER cannot be guaranteed to recover a good $\tau$.

SGA-HMC is motivated by the use of SIMD-hardware to run multiple chains in parallel. Sometimes this is not acceptable due to memory requirements. SGA-HMC can still function when few chains are available, due to cross-iteration averaging courtesy of ADAM and Welford's algorithm, but its convergence speed suffers due to increasing estimator variance. This issue is unlikely to affect long run performance, making the procedure still beneficial compared to NUTS.

SNAPER-HMC is particularly well suited to the Adaptive MCMC framework where $M$ is also learned; empirically, we observed that learning a good $M$ is difficult enough that SNAPER will introduce minimal overhead on top that. If $M$ is fixed, learning a $\tau$ via SNAPER-HMC may be less attractive. This is likely a rare use case, which means that SNAPER-HMC is a good baseline MCMC method to use with SIMD hardware.

\acknowledgments{
We thank Rif A. Saurous for helpful comments on the manuscript and Dave Moore, Colin Carroll and Sharad Vikram for proofreading the implementation.
}

\clearpage
\bibliography{main}

\newpage

\begin{appendices}

\onecolumn

\crefalias{section}{appsection}

\section{Additional Plots}
\label{sec:extra_plots}

\begin{figure*}[!hb]
  \centering
    \includegraphics[scale=0.55]{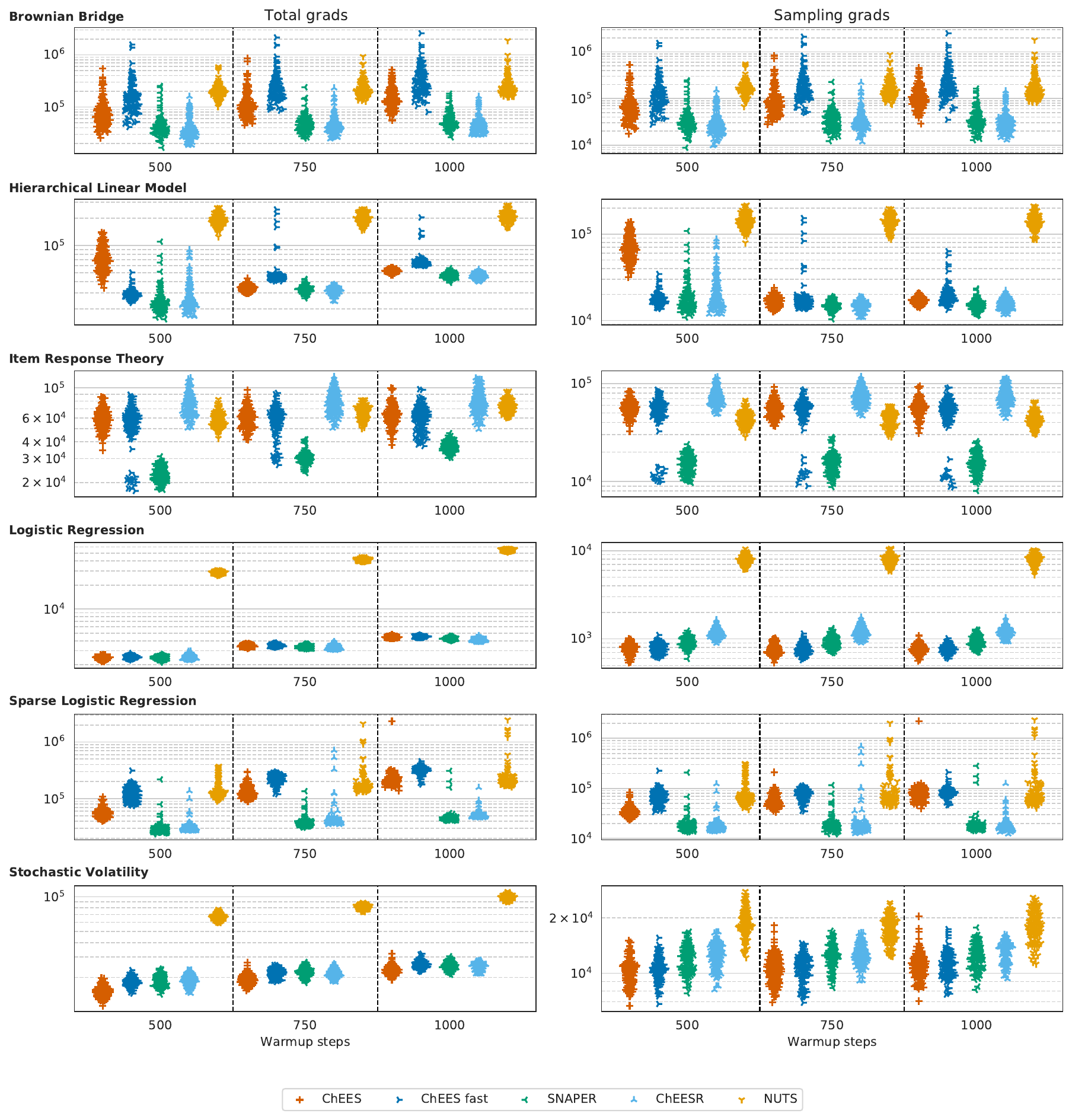}
\caption{
    Gradient evaluation counts as a function of number of warmup steps. See \Cref{fig:early_stop} for details.
}
    \label{fig:early_stop_all}
\end{figure*}

We include two additional plots for the short and long run performance of the considered algorithms, \Cref{fig:asymptotics_all} and \Cref{fig:early_stop_all}.

\begin{minipage}{\textwidth}
    \centering
    \includegraphics[scale=0.55]{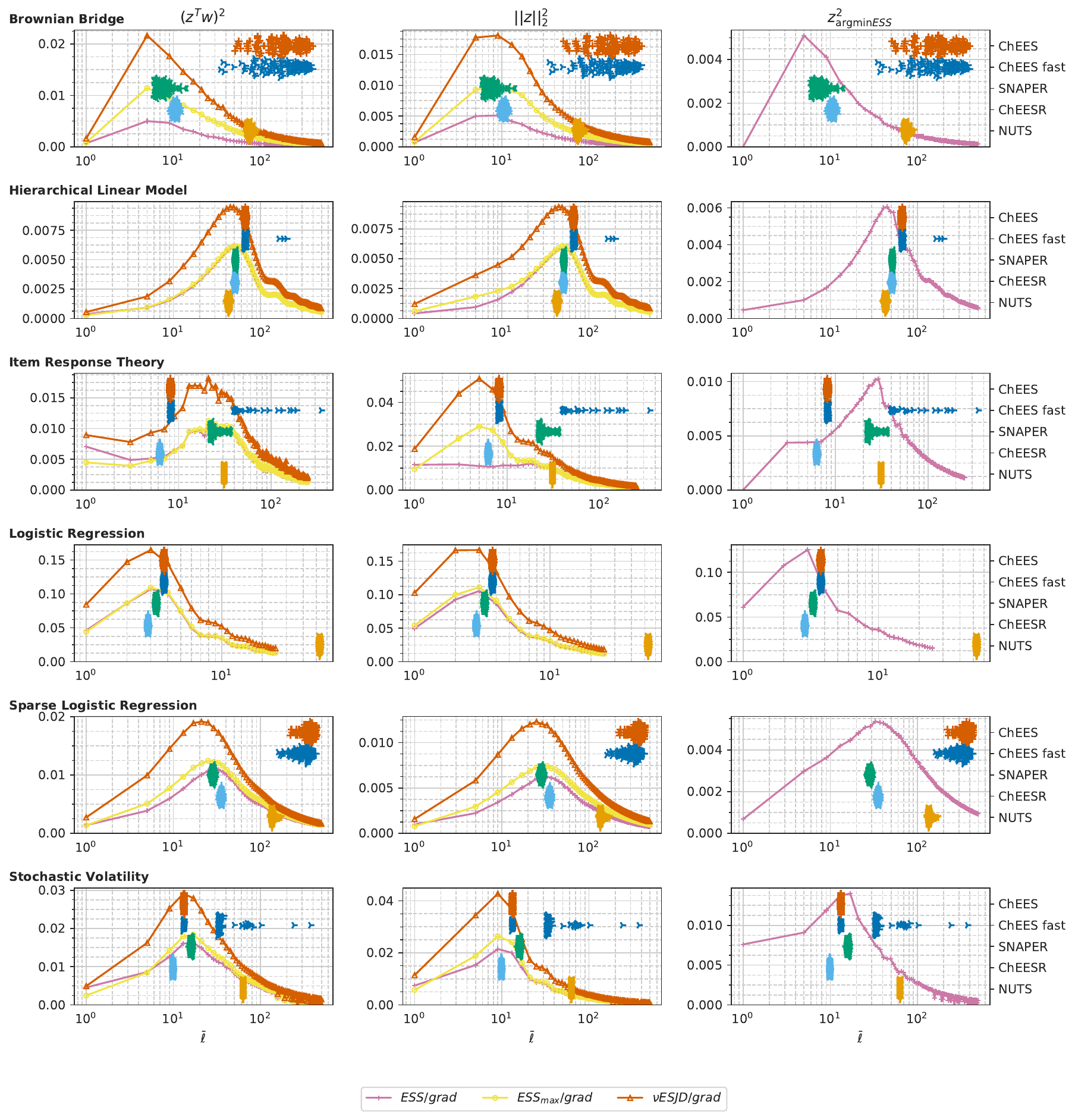}
    \captionof{figure}{Long run behavior of different criteria. See \Cref{fig:asymptotics} for details.}
    \label{fig:asymptotics_all}
\end{minipage}

\clearpage

\begin{table*}[!tb]

\begin{tabular}{ lllllll }
\toprule
 & Grid & ChEES & ChEES fast & SNAPER & ChEESR & NUTS \\
\midrule
Brownian Bridge & 1.018 & 1.025 & 1.023 & 1.019 & 1.023 & 1.020 \\
Hierarchical Linear Model & 1.000 & 1.000 & 1.000 & 1.000 & 1.000 & 1.002 \\
Item Response Theory & 1.004 & 1.002 & 1.002 & 1.000 & 1.004 & 1.000 \\
Logistic Regression & 1.003 & 1.001 & 1.001 & 1.001 & 1.000 & 1.000 \\
Sparse Logistic Regression & 1.011 & 1.001 & 1.001 & 1.010 & 1.018 & 1.026 \\
Stochastic Volatility & 1.005 & 1.005 & 1.004 & 1.004 & 1.007 & 1.003 \\
\bottomrule
\end{tabular}

    \label{fig:asymptotic_rhat}
    \caption{$\hat{R}$ diagnostics corresponding to values in \Cref{tab:asymptotics_tabl}.}

\end{table*}

\section{$ESS^{\max}$ derivation}
\label{sec:ess_max}

We derive the bound $ESS_{f,max} \ge ESS_f$ following \cite{geyer1992practical} and \cite{sokal1997monte}. Given $f(x)$ that satisfies $\int f(x)^2 p(dx) < \infty$ and $x^{(t)}$, $t \ge 0$ forms an ergodic and reversible Markov chain. We can use the spectral theorem to write:
\begin{align}
    \rho_f(l) = \int_{-1}^{1} dE_\lambda \lambda^{|l|},
\end{align}
where $dE_\lambda$ is a probability measure supported on $[-1, 1]$. We can then write:
\begin{align}
    \sum_{l=-\infty}^{\infty} \rho_f(l) &= \int_{-1}^{1} dE_\lambda \, \frac{1 + \lambda}{1 - \lambda}  \\
     &\ge \frac{1 + \int_{-1}^{1} dE_\lambda \, \lambda }{1 - \int_{-1}^{1} dE_\lambda \, \lambda } \\
     \label{eq:bound}
     &\ge \frac{1 + \rho_f(1) }{1 - \rho_f(1) }.
\end{align}
Since $\rho(l) = \rho(-l)$ we have:
\begin{align}
    ESS_f &\triangleq \frac{n}{1 + 2 \sum_{l=1}^{\infty} \rho_f(l)} \\
          &= \frac{n}{\sum_{l=-\infty}^{\infty} \rho_f(l)} \\
          &\le n \frac{1 - \rho_f(1)}{1 + \rho_f(1)} \triangleq ESS^{\max}_f.
\end{align}

\section{Additional SGA-HMC details}
\label{sec:sga_hmc_details}

\begin{algorithm}
\caption{One iteration of Oja's Algorithm with minibatch support}\label{alg:oja}
\begin{algorithmic}[1]
    \Require $x^{(t, k)}$, a minibatch of $d$-dimensional states indexed by batch index $k \in [0, b]$. $w^{(t)}$ current value of the first principal component. $\eta^{(t)}_w$, current learning rate.
    \Ensure $w^{(t + 1)}$, updated value of the principal component.
    \State $\tilde{w} \leftarrow {0}_d$
    \For{$k=1$ to $b$}
        \State $\tilde{w} \leftarrow \tilde{w} + x^{(t, k)} x^{(t, k) T} w^{(t)}$
    \EndFor
    \State $w' \leftarrow \frac{\tilde{w}}{\lnorm \tilde{w} \rnorm}$
    \State $w^{(t + 1)} \leftarrow \frac{w^{(t)} + \eta^{(t)}_w w'}{\lnorm w^{(t)} + \eta^{(t)}_w w' \rnorm}$
\end{algorithmic}
\end{algorithm}

In this section we describe SGA-HMC in more detail. To adapt $\tau$ we use automatic differentiation to compute the pathwise gradients of a criterion. To avoid differentiating the HMC proposal, we manually derive and annotate this gradient. For the criteria we wish to test, we only require the gradients flowing through the proposed\footnote{Technically \Cref{eq:esjd} uses the accepted MCMC state, but we assume the acceptance rate is independent of $\tau$.} state $x'^{(t + 1)}$. Specifically, the vector-Jacobian product of the gradients flowing through $x'^{(t + 1)}$ are:
\begin{align}
    \nabla^T J(x'^{(t + 1)}(\tau)) = \epsilon \sum_{i=1}^d \nabla_i \frac{\partial H(x, p)}{\partial p_i}.
\end{align}
We ignore the fact that for a given $\ell$, $\tau$ can only take discrete values due to the discretized dynamics used to implement HMC. Now, given $\tau \sim \unif(0, 2 \bar{\tau})$ and a differentiable criterion we are thus able to compute $\frac{d \calC(\tau)}{d \tau}$. We use ADAM to adapt $\bar{\tau}$. When learning $\epsilon$ we  use a synthetic gradient to push the harmonic mean of per-chain acceptance probability $\alpha^{(b)}$ towards a target acceptance rate $\alpha^* = 0.8$:
\begin{align}
    \widehat{\frac{\partial\mathcal{A}}{\partial\alpha}} = \alpha^* - \left(\frac{1}{b} \sum_k \alpha_k^{-1}\right)^{-1}.
\end{align}
A harmonic mean is used to make sure that all chains that are evaluated in parallel and share HMC hyperparameters can make progress \citep{hoffman2021adaptive}. $\epsilon$ is also adapted using ADAM. When HMC is parameterized by $\tau$ rather than $\ell$ it is important to initialize $\epsilon$ to a reasonable value so as not to waste a lot of compute in the early iterations. Standard protocol is to do a search over $\epsilon$ until a reasonable acceptance probability is obtained \citep{hoffman2011no}. To make more efficient use of gradient evaluations, we instead force HMC to only take a single leapfrog step for the first 100 iterations. This procedure recruits ADAM to perform the search for us, while also carefully navigating the potentially pathological geometry if the algorithm is initialized in a suboptimal location.

The mean $\mu$ and variance $\sigma^2$ of the state are learned using a variant of Welford's algorithm \citep{welford1962note} which uses an exponentially weighted average with a decaying learning rate to smoothly discount early, non-representative iterates. To aid trajectory adaptation, instead of using raw $\sigma^2$ as the value for the diagonal of $M$, we normalize it by the largest component. This has the effect of minimizing the effect of learning $M$ on the optimal value for $\bar{\tau}$ (which ideally depends primarily on the longest length scale of the distribution). ChEES, ChEESR and SNAPER require (see \Cref{eq:snaper,eq:f_chees}) $x$ to be centered. This center is computed by applying the same algorithm as is used to learn $\mu$ except the update is derived either from the current state $x^{(t)}$ or acceptance-probability weighted proposed state $\frac{1}{\sum_k \alpha_k} \sum_k \alpha_k x'^{(t + 1, k)}$ when computing $z^{(t)}$ and $z'^{(t + 1)}$ respectively.

\section{Hyperparameter settings}
\label{sec:hyperparameters}

The hyperparameters were minimally tuned, and shared across all problems.

\begin{minipage}{\textwidth}
\centering
\begin{tabular}{ lll }
\toprule
Hyperparameter & Meaning & Value \\
\midrule
$b$ & Number of chains ran in parallel & 64 \\ 
$\kappa$ & Factor controlling moment adaptation & 8 \\
$\eta_{\bar{\tau}}$ & Learning rate for $\bar{\tau}$ & 0.05, 0.025 (ChEES) \\
$\eta_w^{(t)}$ & Learning rate for $w$ & $\frac{8}{t}$ \\
$\eta_\epsilon$ & Learning rate for $\epsilon$ & 0.05 \\
$\beta_1$ & ADAM hyperparameter for $\bar{\tau}$ & 0 \\
$\beta_2$ & ADAM hyperparameter for $\bar{\tau}$ & 0.5 (short run), 0.95 (long run)\\
\bottomrule
\end{tabular}
\end{minipage}

\end{appendices}

\end{document}